# A Four-channel Optically Pumped Magnetometer for a Magnetoencephalography Sensor Array

JOONAS IIVANAINEN,[1] TONY R. CARTER,[1] JONATHAN E. DHOMBRIDGE,[1,2] TIMOTHY S. READ,[1,2] KALEB CAMPBELL,[1,2] QUINN ABATE,[1] DAVID M. RIDLEY[1,2], AMIR BORNA,[1] AND PETER D. D. SCHWINDT[1,*]

[1]*Sandia National Laboratories, Albuquerque, NM 87123, USA*
[2]*Center for Quantum Information and Control, Department of Physics & Astronomy, University of New Mexico, Albuquerque, NM 87106, USA*
*\*pschwin@sandia.gov*

**Abstract:** We present a novel four-channel OPM sensor for magnetoencephalography that utilizes a two-color pump/probe scheme on a single optical axis. We characterize its performance across 18 built sensor modules. The new sensor implements several improvements over our previously developed sensor including lower vapor-cell operating temperature, improved probe-light detection optics, and reduced optical power requirements. The sensor also has new electromagnetic field coils on the sensor head which are designed using stream-function-based current optimization. We detail the coil design methodology and present experimental characterization of the coil performance. The magnetic sensitivity of the sensor is on average 12.3 fT/rt-Hz across the 18 modules while the average gradiometrically inferred sensitivity is about 6.0 fT/rt-Hz. The sensor 3-dB bandwidth is 100 Hz on average. The on-sensor coil performance is in good agreement with the simulations.

## 1. Introduction

Optically pumped magnetometers (OPMs; Ref. [1]) operating in the spin-exchange-relaxation-free (SERF) regime [2] are among the most sensitive magnetic-field sensors, achieving sensitivity levels as low as 0.16 fT/rt-Hz [3] and 0.54 fT/rt-Hz [4]. OPMs are based on the interaction of spin-polarized atoms with magnetic fields. Optical pumping is used to transfer the angular momentum of a pump beam to atoms (typically in an alkali metal vapor) spin-polarizing them. The spin polarization interacts with the magnetic field through Larmor precession, and by optically detecting the projection of the spin polarization along the probe beam propagation direction, the external magnetic field can be determined. In the SERF regime, characterized by high atom density and near zero magnetic field, the OPM sensitivity can be greatly enhanced as the spin polarization relaxation due to spin-exchange collisions is strongly suppressed.

Recently, there has been an increasing interest in applying OPMs in biomagnetism, i.e., the measurement of magnetic fields from the human body. OPMs have been used to measure the magnetic fields of the human brain [5-11], the heart [12-14], the peripheral nerves [15], muscles [16, 17], the retina [18], and the spinal cord [19]. One particularly interesting application of OPMs is in magnetoencephalography (MEG).

MEG is a neuroimaging technique in which the magnetic fields of the human brain are detected outside the head [20]. Historically, superconducting quantum interference devices (SQUIDs) have been used to detect MEG signals. However, the use of cryogenics necessitated by liquid helium cooling of the SQUIDs brings several disadvantages to SQUID-based MEG systems. First, the thermal insulation necessitated between the subject and the SQUIDs introduces a gap of about 2 cm between the SQUID sensors and the subject's scalp limiting the spatial resolution and information of the MEG measurement [21, 22]. Second, the one-size-fits-all SQUID helmet is not adjustable to the subject's head shape further limiting the spatial resolution, especially in children. Third, the heavy and bulky SQUID-MEG systems do not allow scanning of moving and ambulating subjects beyond the limitations of the rigid helmet.

To this end, OPMs have been introduced as an on-scalp MEG sensor detecting the magnetic field within approximately 5 mm from the scalp and, potentially, enabling higher spatial resolution than that obtained by SQUIDs [22, 23]. Due to their compact size and small weight, OPMs may also be used to build wearable arrays that enable scanning of moving subjects [24]. To this date, numerous OPM sensor designs have been introduced to MEG applications. The majority of the OPM sensors are based on alkali metal atoms such as rubidium-87 ($^{87}$Rb; [25-27]), potassium [7, 28] and cesium [29, 30] while some sensors use helium-4 [31]. The magnetic sensitivities of the $^{87}$Rb-OPMs designed for MEG applications have been about 10 fT/rt-Hz with bandwidths around 100 Hz. The helium-4 OPM can achieve higher bandwidth of about 2 kHz with a magnetic sensitivity of roughly 50 fT/rt-Hz.

In this paper, we describe our new and improved four-channel SERF-OPM sensor that is based on our $^{87}$Rb-OPM utilizing a two-color pump/probe scheme on a single optical axis [25, 32, 33]. We have implemented various changes to the sensor lowering its operating temperature and optical power requirements. We have also designed and implemented new electromagnetic field coils onto the sensor head that provide tri-axial control of the magnetic field at the channel positions. We will present the detailed performance of a single sensor and the performances of 18 sensor modules built so far. We also outline the on-sensor coil design methodology using stream-function-based current optimization [34] and present experimental characterization of the coils.

## 2. Overview of the OPM sensor

In this section, we present an overview of our next-generation OPM sensor. The sensor is based on our previously published sensor described in Ref. [25] utilizing a two-color pump/probe scheme on a single optical axis. The key differences are highlighted as relevant below. We use circularly polarized 795-nm light to optically pump $^{87}$Rb atoms inside a vapor cell. As a result of optical pumping, spin polarization is created among the $^{87}$Rb atoms that interacts with magnetic field through Larmor precession. We detect a component of the spin polarization along the pump beam by measuring the angle of a linearly polarized probe beam at 780 nm after it has passed through the atomic gas and interacted with the atoms via Faraday rotation. The collinear pump and probe beams are tuned close to the D1 and D2 transitions of $^{87}$Rb, respectively. In the measurements described in this paper, the pump and probe light are generated by distributed feedback lasers.

The sensor is operated in the SERF regime near zero magnetic field. We utilize magnetic field modulation at 1 kHz along an axis transverse to the light beams to reduce 1/$f$-noise, to define the sensitive axis, and to linearize the sensor response to a magnetic field [35, 36]. By utilizing lock-in detection at 1 kHz, the optical Faraday rotation signal carried by the probe beam has a linear response to external transverse-field changes.

Figure 1A gives a schematic of the sensor optics. A polarization-maintaining optical fiber brings both 795-nm (pump) and 780-nm (probe) light into the sensor module. The co-propagating pump and probe beams then go through a polarizer that ensures that both beams are linearly polarized. After the polarizer, the beams pass through a dual-wavelength waveplate that acts as half- and quarter-wave plates for the probe and pump beams, respectively. After the waveplate, the probe beam maintains linear polarization while the pump beam's linear polarization is converted to a circular polarization. The beams are then collimated with a lens and go through a diffractive optical element that splits both beams into four equally spaced beams that travel to the vapor cell where they interact with $^{87}$Rb atoms (Fig. 1B).

Similar to our previous design, the beams are separated by 18 mm within the vapor cell and travel 4 mm through the cell before they are reflected back from the outer back side of the vapor cell. In contrast to our previous sensor, the full-width-half-maximum (FWHM) diameters of the pump and probe beam are designed to be ~2 mm; after alignment they vary from 1.75 to 2 mm. The paths of the pump and probe beams in the vapor cell define the positions of the four magnetometer channels of the sensor.

The vapor cell has a 24 mm × 24 mm × 4 mm internal volume and has an antireflection coating on the laser input side of the cell and a high reflectivity coating on the opposite side of the cell. The cell has no internal optical coatings. It is filled with a small amount of isotopically enriched rubidium-87 and a nitrogen buffer gas. We have reduced the pressure of the nitrogen buffer gas from 600 Torr in our previous sensor to 300 Torr in our new sensor. This allows us to operate the sensor at a lower temperature as we can reduce the $^{87}$Rb atom density by a factor of two while maintaining a similar optical depth.

To ensure sufficient $^{87}$Rb number density, the vapor cell is heated with a pair of ceramic heaters (Thick Film Technologies) on the top and bottom surfaces of the vapor cell. An AC current at roughly 200 kHz is driven through the electrical traces in the heaters. Driving the heater at a high frequency minimizes the interference near the 1-kHz modulation. The two heaters could in principle be driven with currents with different amplitudes; so far, we have achieved satisfactory performance by driving them with equal voltages. We monitor the temperature of the vapor cell by measuring the resistance of separate high-resistance (~600 Ω) electrical traces on both top and bottom ceramic heaters. Using free-induction decay measurements of the linewidth, we measure the atom density from which we estimate the vapor cell temperature in the current sensor to be ~135 °C. This is a considerable reduction from the 150 °C operating temperature of the previous sensor. To reduce the external surface temperature of the sensor module, the vapor cell is surrounded by a 5 mm aerogel insulation blanket (Cryogel from Aspen Aerogels, Inc.). The improved insulation eliminates the need for air cooling used in the previous sensor. The distance from the center of the cell to the sensor bottom surface touching the subject's head is 9 mm.

After traversing through the vapor cell, the probe beams from the four channels are detected using balanced polarimetry with revised detection optics. Compared to the previous design, the new detection optics collect the probe light using an individual detection module for each channel (see Fig. 1B). Each beam reflected from the vapor cell first passes through an interference filter with a pass-band at 780 nm to filter out pump light. Then the probe light is split into two beams with a polarizing beam splitter (PBS) for measurement of the polarization angle with two photodiodes. Before the PBS, a 780-nm half-wave plate is used to balance the probe light intensity between the two arms of the photodetector. The photodiode difference signal is then brought out from the sensor module and fed to a transimpedance amplifier.

The new detection optics design improves upon the old design in several ways. First, the modularity of the detection optics allows for control of the half-wave plate angle for each of the four channels to balance the two probe-beam arms for maximum common-mode rejection. Second, the design optimizes the PBS performance as the beam enters it at the design angle. Third, in the previous design, the photodiodes for each channel were placed very close to each other, and substantial optical crosstalk between channels was observed. The new design eliminates the crosstalk with widely separated detection modules while easing alignment tolerances for faster optical assembly.

As with our previous sensor, we use two separate pump beams detuned by ±10 GHz around the zero light-shift point [25]. By changing the power balance between the pump beams, the net light shift can be zeroed while maximizing sensitivity. We have found that the optimal pump power ratio of the two beams is around 28% with the positively-tuned component having more power (see also Ref. [25]). The probe beam is detuned by about -70 GHz from the D2-transition of $^{87}$Rb.

Fig. 1C shows a CAD drawing of the sensor together with the optical paths of the light beams, while Fig. 1D shows a picture of a built sensor module. Altogether, the footprint of the sensor on the subject's head is 40 mm × 40 mm while the sensor length is 26.6 cm.

Electromagnetic coils wrapped around the sensor head (Fig. 1D) provide field modulation at 1 kHz with an amplitude of roughly 140 nT and three-axis field control at the four channel locations. In the following sections, we will detail the coil design methods and performance, in addition to the overall performance of the sensor.

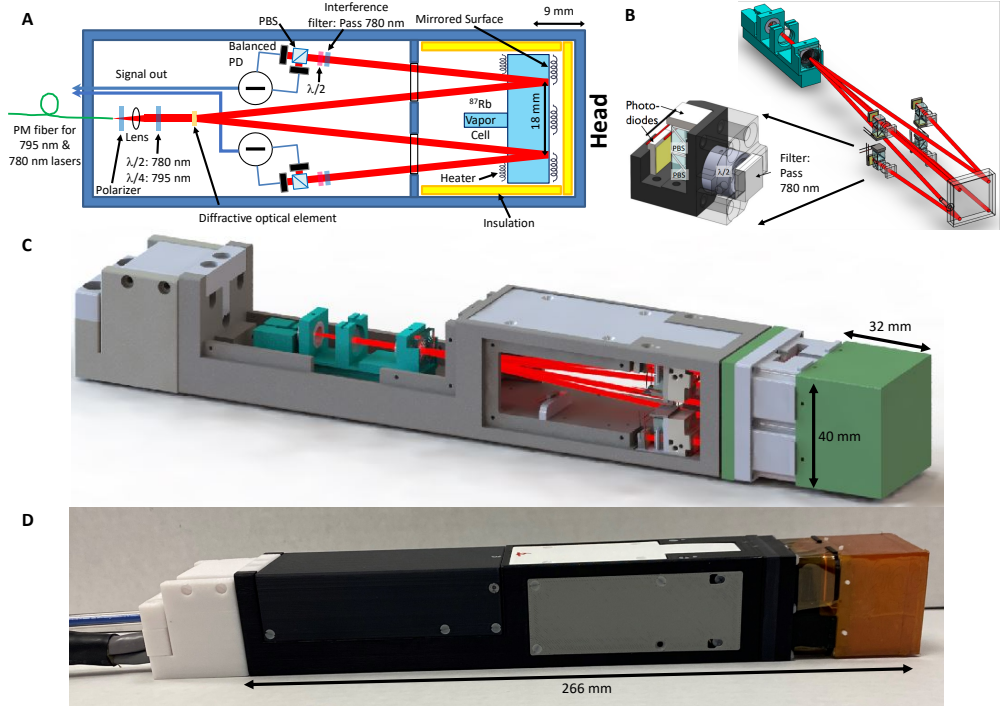

**Figure 1:** The next-generation OPM sensor. **A:** A schematic of the sensor module showing its components. **B:** A CAD drawing of the optics components and the optical paths of the light beams inside the sensor. The inset shows the detection optics, used to measure the Faraday rotation angle of the probe light, in detail. A CAD drawing (**C**) and a picture (**D**) of the sensor.

## 3. On-sensor coil design and experimental characterization

In this section, we present the novel on-sensor coil design of the sensor. Compared to the on-sensor coil design of our previous sensor, the new design improves the field homogeneity as well as orthogonality of the field components at the channel locations. Field homogeneity is important to minimize gradients in the channel volumes that might degrade sensitivity of the channels. The new design also implements individual control of the longitudinal field at each channel; these field components are important to null as they may cause cross-axis projection error (CAPE; Ref. [37]). We give detailed descriptions of the coil-design method and the experiments we performed to characterize the performance of the coils.

### 3.1 *Coil design and characterization methods*

We want to control the magnetic field along the transverse axes ($B_x$ and $B_y$) and the longitudinal optical axis ($B_z$) at the four channel locations with the on-sensor coils. The transverse on-sensor coils also provide field modulation defining the sensitive axis of the OPM. For the transverse field control/modulation, we designed a pair of $B_x$-coils on the outside surface of the sensor producing a homogeneous transverse field at all four channel locations. The $B_y$-coil is identical to the $B_x$-coil.

As longitudinal field offsets cause CAPE in the OPM, we wanted to have individual control of the $B_z$ field at each channel. We designed a $B_z$-coil onto the OPM bottom surface closest to the subject's head that provides a homogeneous $B_z$ field for a single channel with minimal leakage to the other channels of the sensor. The $B_z$-coil was constrained to cover 1/4$^{th}$ of the total surface area of the bottom of the sensor. By placing four non-overlapping $B_z$-coils on the surface, individual $B_z$-field control can be provided to all the sensor channels.

The current paths for the on-sensor coils were designed using the numerical stream-function-based target field method implemented in the *bfieldtools* Python software package [38, 39]. The method uses quadratic programming to optimize a current on a set of surfaces (represented as triangle meshes) to produce a desired magnetic field at a set of target points. Additionally, it is possible to simultaneously minimize the stray field at a set of stray-field points. The software minimizes either the resistive power or the magnetic energy of the surface current with constraints on the field at the target and stray points. After the optimal surface current is found, it is discretized to current loops at its contours.

The geometry used to optimize the on-sensor coils of the OPM is presented in Fig. 2. The coil surfaces for the $B_x$-coil were modeled as a 38 x 25 mm$^2$ triangular meshes (2,597 vertices; 5,200 triangles) separated by 4 cm. The $B_z$-coil mesh was an 18 x18 mm$^2$ square (3,028 vertices; 5,854 triangles). Each channel location (target region) was modeled as a cylindrical volume (diameter: 2.5 mm; length: 4 mm) and was discretized to 15 points having approximately uniform coverage of the volume. The resistive energy of the $B_x$-coil was minimized with the constraints that the field be homogeneous at all channel locations to an absolute tolerance of 10%, and the stray field at points on a cylindrical shell of radius 5 cm around the sensor center be less than 12% of the field magnitude at the channel locations. Before optimization, the surface current for the $B_x$-coil was truncated to the first 30 surface harmonics [38]. For the $B_z$-coil, we minimized its magnetic energy with a constraint that it produces homogeneous field along $z$ at a single channel location (absolute tolerance: 10%). We also ensured that its stray field would be less than 20% at points on a cylindrical shell of radius 1.5 cm around the channel location.

In addition to the optimization of the field shape, coil efficiency has to be in a correct range. The field modulation requires the $B_x$- and $B_y$-coils to generate several hundreds of nT using a current of roughly 20 mA. The efficiency of the $B_z$-coil can be lower; the $B_z$-coils are used to null DC fields of tens of nT. These requirements were monitored when the coil loops were designed.

After finding the optimal current distribution, the coil loops for the $B_x$- and $B_z$-coils that produced the desired combination of field homogeneity and efficiency were searched by iterating over the number of contours. For this task, we calculated the field at a set of high-resolution evaluation points at the channel locations (203 points per channel). Additional smoothing of the loops and removal of loops with lengths less than 5 mm was performed for the $B_z$-coil.

The optimized coil loops were exported to the KiCad software (*https://www.kicad.org/*) and the coils were implemented using a two-layer flexible printed circuit board (PCB). The same coil loops were printed to both sides of the two-layer PCB. Electrical vias connecting the two sides of the PCB were used to "spiralize" the coil loops to produce a single conductor.

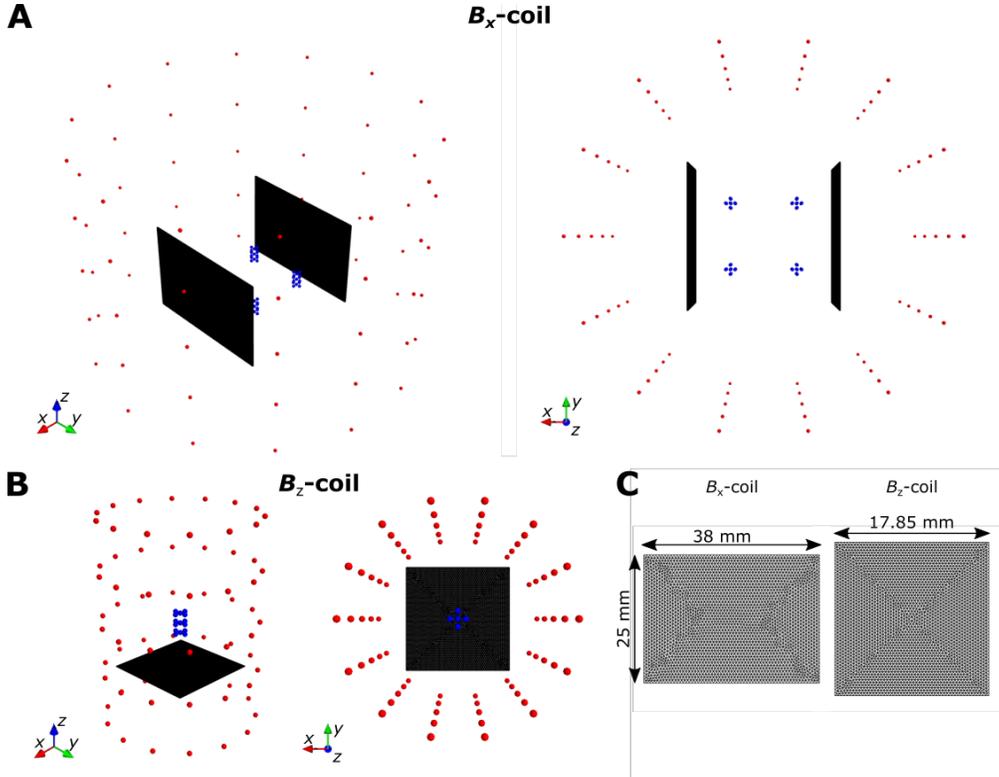

**Figure 2:** The geometries for optimization of transverse ($B_x$; panel **A**) and longitudinal ($B_z$; panel **B**) on-sensor coils of the sensors. In panels **A** and **B**, the target and stray field points are presented with blue and red dots, respectively, while the optimization surfaces for the current are shown as black meshes. **C:** Close-ups of triangle meshes for $B_x$- and $B_z$-coils. Please note that the meshes are not to scale but magnified to show detail.

*Field efficiency and orientation measurements*
We performed the following routine to measure the efficiencies (nT/mA) of the on-sensor coils and the orientations of the generated fields at the channel locations after installation of the flex-PCB coil set on the sensor head. The sensor was placed inside a tabletop cylindrical magnetic shield (MS-2, Twinleaf LLC, Plainsboro Township, NJ, USA), the on-sensor coil was driven with a known DC current, and linear sweeps of the magnetic field along all three orthogonal directions were performed separately by using the approximately orthogonal field coils of the shield with known calibration. The raw photodiode outputs were sampled with a data-acquisition (DAQ) unit (NI USB-6289, National Instruments, Austin, TX, USA). The sensor output to the scan is a Lorentzian function [40] whose peak location gives the magnetic field offset along the axis of the field scan. The scans were performed along the three orthogonal axes for each on-sensor coil at a single current setting. The background magnetic field inside the shield was measured and subtracted from the results. To obtain the Lorentzian lineshape when scanning along the z-axis, a DC field along the y-axis was applied. A Lorentzian was fit to the measured data and its peak location was extracted using MATLAB.

The sensor was heated to the approximate operating temperature. Only a single pump laser was used that was positively detuned as described in Sec. 2. Both pump and probe power were reduced to a level around 0.01 mW using a neutral density (ND) filter to reduce fictitious magnetic fields due to light shifts [41].

*Field inhomogeneity measurements*

Two different approaches were used to measure the inhomogeneities of the installed on-sensor coils at the channel locations. For the transverse coils ($B_x$ and $B_y$), we performed free-induction decay (FID) measurements to characterize the $T_2$ gradient broadening caused by the applied fields [42]. The sensor was operated at a lower temperature than in the magnetometer operation to ensure that the sensor was operating outside the SERF regime. We also used a single positively-tuned pump laser operated at a higher power (12 mW) and switched at 129 Hz with an optical chopper (SR 540, Stanford Research Systems, Sunnyvale, CA, USA). We operated the sensor this way to ensure the atoms were fully polarized throughout the whole channel volume. The transverse field was turned on while the pump was blocked and the resulting FID signals were measured.

An exponentially decaying sinusoid was fit to the measured FID signals using MATLAB and the precession frequency and $T_2$ were extracted from the fit. Similar to Tayler et al. [42], we quantified the field inhomogeneity as the derivative $\partial \Delta\vartheta / \partial f_L$ where $\Delta\vartheta$ is the FWHM given by $T_2$ and $f_L$ is the Larmor frequency. By comparing the measured precession frequency to the estimated Larmor frequency $f_L = \gamma B$, where $\gamma/2\pi = 7$ Hz/nT is the gyromagnetic ratio for $^{87}$Rb, we can also estimate the field generated by the coils and determine corrections to the coil efficiencies measured as described in the previous section.

To estimate the $B_z$-coil field inhomogeneity, we measured the gradient broadening of the relaxation time by running the sensor in the $M_z$-mode [43] as follows. The sensor was heated to a temperature lower than the magnetometer operating temperature. Only a single positively-tuned pump laser was used, and the pump and probe power were reduced to a level of around 0.01 mW. The frequency of sinusoidal modulation of the magnetic field along the *y*-axis was swept from 50 Hz to 12 kHz at a rate of 1 Hz; the peak-to-peak amplitude of the magnetic field was roughly 40 nT. We then applied DC currents to the on-sensor $B_z$-coils. The output of the sensor to the $B_y$ frequency sweep is a Lorentzian; we can extract its peak location and FWHM to estimate the generated field and its inhomogeneity due to the resonance linewidth broadening. For comparison, we also performed both $B_x$ and $B_z$ inhomogeneity measurements for the large coils of the shield whose fields should be substantially more homogeneous across each individual channel volume.

### 3.2 On-sensor coil results

Fig. 3A shows the optimized on-sensor coil designs while Fig. 3B shows pictures of the manufactured flexible coil PCBs. In Fig. 3B we also show the electric via design connecting the two layers of the board. Fig. 4 shows the simulated magnetic field maps of the on-sensor coils inside and outside the sensor.

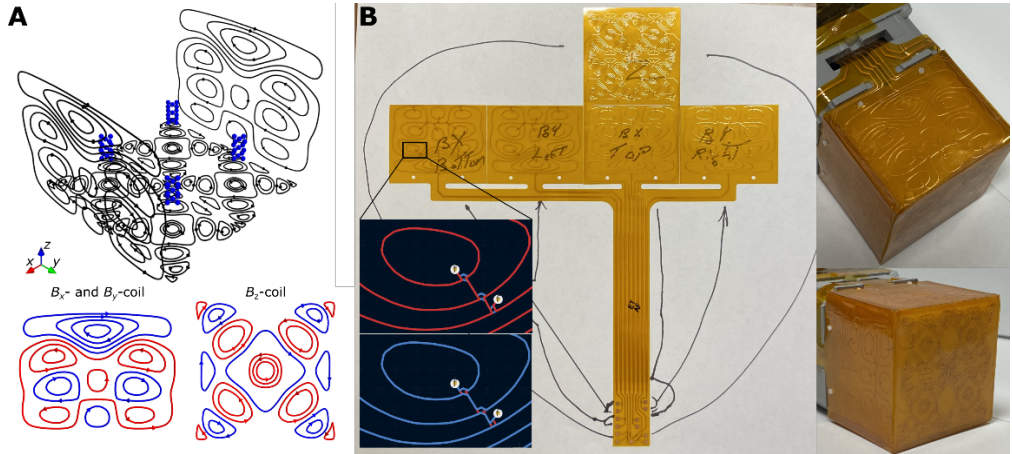

**Figure 3:** The optimized on-sensor coils. **A:** The optimized current loops for the transverse ($B_x$ and $B_y$) and longitudinal ($B_z$) on-sensor coils. The channel locations are indicated with blue dots. In the bottom panel, the red and blue traces indicate the direction of the current (clockwise and counter-clockwise, respectively). **B:** Pictures of the manufactured on-sensor coil two-layer flexible PCB lying flat or wrapped around the sensor head. The inset shows the coil traces on both layers of the flexible PCB (red/blue) and their connections with electric vias to form a single coil.

The simulated design efficiency of the $B_x$-coil is 28.8 nT/mA while its maximum inhomogeneity across the volume of each channel is 11.4%. The magnetic field leakage is around 50%, 15% and 7% at distances of 4.0, 5.0 and 6.0 cm from the center of the coils. The designed efficiency of the $B_z$-coil is 3.4 nT/mA, and its maximum inhomogeneity is 13.6% at a given channel location. The cross-channel leakage of the field is at maximum 16% to the neighboring channel location. The leakage is 12% when the field is averaged over the neighboring channel location.

Figure 5A gives examples of the measured Lorentzian curve and their fits when the field is scanned along the $x$-axis with the shield coil and DC field is applied via the $B_x$ on-sensor coil. The figure also shows the measurement of the $x$-axis component $B_x$ of the background magnetic field in the shield obtained using the same procedure but without applying any current to the on-sensor coil. The difference between the estimated peak positions of the Lorentzians between these two is used to estimate the field generated by the on-sensor coil. Dividing this field by the applied current, the estimated field efficiencies of the $B_x$ on-sensor coil at the four channel positions are measured to be 30.6, 29.6, 26.9 and 27.2 nT/mA for each of the four channels. The corresponding values for the $B_y$ on-sensor coil are 29.2, 28.1, 29.1 and 27.4 nT/mA. The maximum relative errors to the design value (28.8 nT/mA) are thus 6.5% and 4.8% for the $B_x$ and $B_y$ on-sensor coils, respectively. Fig. 5B shows the estimated orientations of the generated $B_x$ and $B_y$ on-sensor coil fields. The fields are orthogonal within 5.4° (average: 3.6°) at the channel positions while the fields lie on the $xy$-plane to within 2.9°.

The same procedure gives the $B_z$ on-sensor coil efficiencies 3.3, 3.4, 3.3 and 3.3 nT/mA at each of the channel positions. The maximal deviation from the design value of 3.4 nT/mA is 3.6%. Maximum $B_z$-coil leakages to neighboring channels are 12.8%, 12.6%, 10.7% and 12.2%. The generated $B_z$-fields are orthogonal to within 5.6° to the $B_x$ and $B_y$ on-sensor coil fields (average: 2.7°). We note that the measured values are affected to some degree by the inhomogeneities and orientation errors of the shield coils sweeping the field.

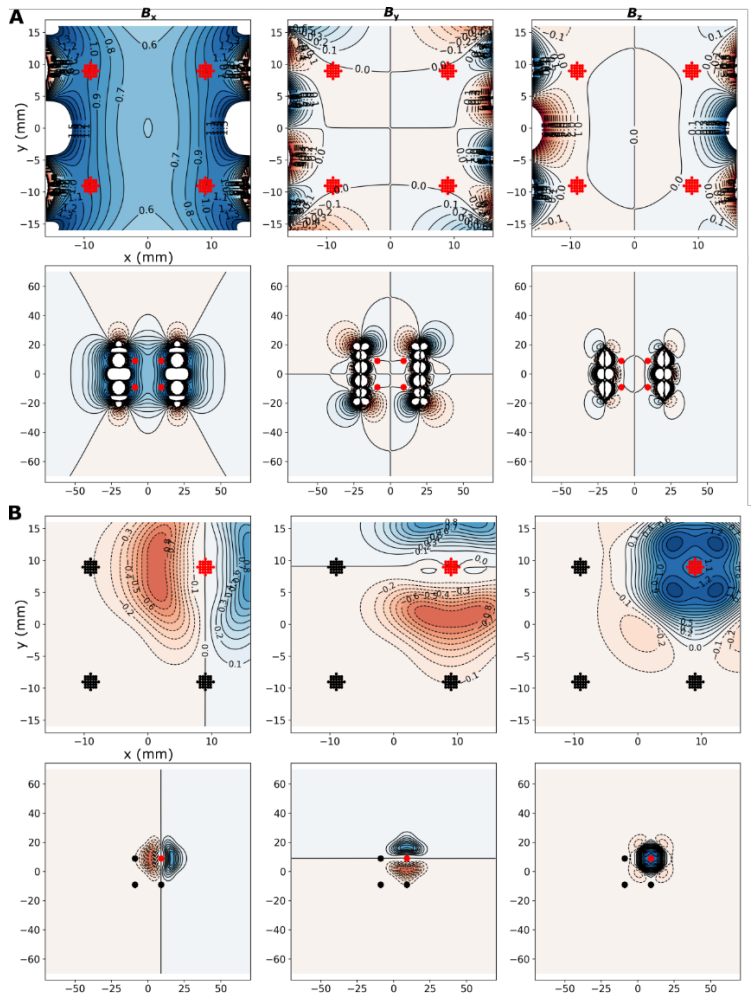

**Figure 4:** The simulated magnetic fields of the transverse ($B_x$; panel **A**) and longitudinal ($B_z$; panel **B**) on-sensor coils inside and outside the sensor. Channel locations are shown with dots; target points are colored red while stray points are colored black. The coil fields are normalized to the average field at the target points and the 10% contours are shown. The $x$-, $y$- and $z$-components of the field are shown in the left, center and right columns, respectively.

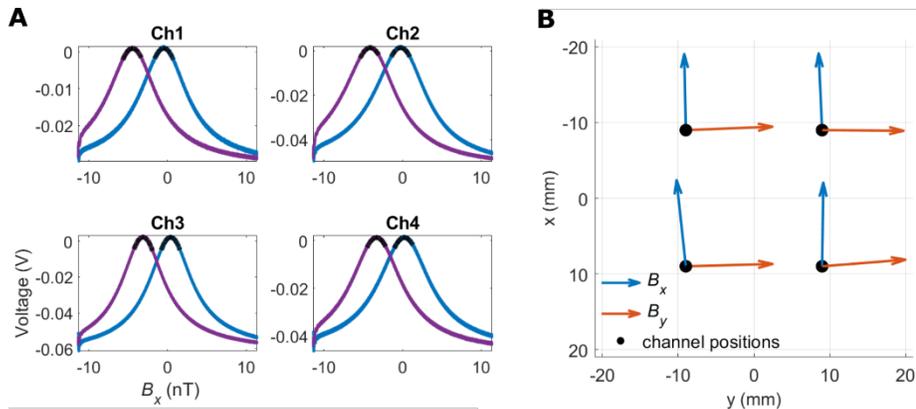

**Figure 5:** Estimating the on-sensor coil efficiencies (nT/mA) and orientations by sweeping the magnetic field along the three orthogonal axes. **A:** The resulting Lorentzian lineshapes for all four channels of the magnetometer when the magnetic field is swept along the $x$-axis. The blue curves show the lineshapes when no magnetic field is applied inside the shield while the magenta curves give the lineshapes when field is applied using the $B_x$ on-sensor coil. **B:** The estimated field orientations of the transverse on-sensor coils at the four channels.

Examples of the measured FID-signals when a transverse on-sensor coil is excited are shown in Fig. 6A with the corresponding exponentially decaying sinusoid fits. Figure 6B shows the estimated linewidths and Larmor frequencies of the FID-signals for a single channel and for both transverse on-sensor coils. The measured linewidths give field inhomogeneity estimates across the four channel volumes of 4.7%, 8.0%, 13.4% and 9.9% for the $B_x$ on-sensor coil and 8.0%, 10.9%, 7.9% and 8.9% for $B_y$, respectively. The Larmor frequency fits give modest corrections ranging from 0.1% to 3.0% to the transverse on-sensor coil efficiencies measured previously (see the paragraph above). The linewidth as a function of the shield coil field shows a small positive slope of 0.9% at maximum across the channel locations.

Fig. 6C shows examples of the $B_y$ frequency sweeps when the sensor is run in the $M_z$-mode, and either the $B_z$ on-sensor or shield coil is excited with DC currents. The gradient broadening is easily visible in the $B_z$ on-sensor coil measurement. The fitted linewidths in Fig. 6D give $B_z$ on-sensor coil field inhomogeneity estimates of 22.3%, 22.6%, 19.2% and 19.4% across the four channel volumes, respectively. The slope of the linewidth–Larmor frequency curve when the shield is coil exited is approximately zero on average across the channels. The field efficiencies for the $B_z$ on-sensor coils given by the Larmor frequency fits are 3.5, 3.4, 3.4 and 3.6 nT/mA, respectively.

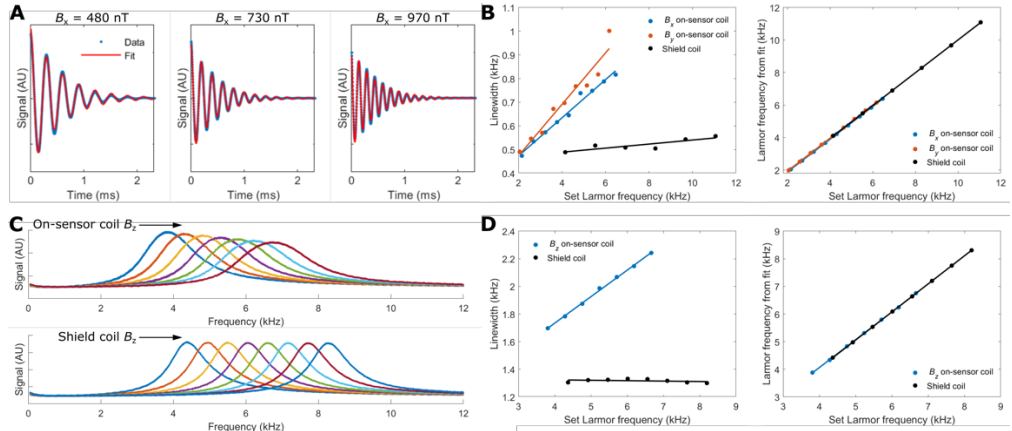

**Figure 6:** The measured inhomogeneities and efficiencies of the on-sensor coils. **A**: Example free-induction-decay (FID) signals when a transverse ($B_x$) on-sensor coil is excited. The data are shown with blue dots while the exponentially-decaying-sinusoid fit is shown with a red line. **B**: The estimated linewidths and precession frequencies of the FID signals of a single magnetometer channel when a transverse on-sensor coil ($B_x$ or $B_y$) or a large shield coil is supplied using increasing currents. **C**: Example outputs to a $B_y$ frequency sweep when the sensor is run in the $M_z$-mode and either the on-sensor (top) or the shield $B_z$ coil (bottom) is excited with DC currents. Each colored curve shows the response to a single $B_y$ frequency sweep. The successive curves correspond to increasing DC fields generated by the coils. The broadening of the resonance as a function of DC field amplitude is well visible in the on-sensor coil data. **D**: The estimated linewidths and peak frequencies of the Lorentzian curves of a single channel shown in panel C. The slopes of the Larmor frequency–linewidth plots can be used to estimate the inhomogeneity of the applied field within the channel volume. The Larmor frequencies shown on the plot $x$-axes are given by the applied currents and field efficiencies estimated from the data shown in Figure 5. The $y$-axis Larmor frequencies are estimated from the signals and their fits shown in panels A and C; the slopes of these plots give corrections to field efficiencies measured in Fig. 5.

## 4. Sensor performance

In this section, we describe the measured performance of a sensor module and the related experimental methods. We describe its operating parameters as well as bandwidth and sensitivity of its four channels. We analyze the contributions of different noise sources in detail to show the factors that limit the sensor sensitivity. We also report the performance statistics across the 18 sensor modules we have built so far.

### 4.1 *Sensor performance evaluation methods*

To quantify the performance of the sensor, we measured the magnetic sensitivities and bandwidths of the four sensor channels. We also operated the sensor as a synthetic gradiometer by digitally subtracting the outputs of the adjacent channels; we report the common-mode rejection ratios (CMRRs) and the sensitivities of the four gradiometer channels. We also give the histograms of laser powers, sensitivities and bandwidths for each of the 18 sensor modules.

The measurements were performed inside the tabletop cylindrical magnetic shield (MS-2, Twinleaf LLC). To reduce magnetic noise inside the shield, we placed a custom cylindrical ferrite layer into the shield [44]; the ferrite layer was the innermost layer of the shield. The sensor was operated with a custom three-axis coil set inside the ferrite shield layer; the shield coils were calibrated using a fluxgate magnetometer (FGM3D, Sensys GmbH, Bad Saarow, Germany). The field was modulated at 1 kHz along an axis transverse to the sensor optical axis ($z$) using the shield coils. The 1-kHz modulation was generated using a lock-in amplifier (SR830, Stanford Research Systems).

The raw photodiode outputs were sampled at 100 kHz with a DAQ unit (NI USB-6289). Software lock-in detection was used for demodulation; the demodulated (magnetometer) signals were down-sampled to 1 kHz by first applying a fourth-order RC-filter with a time constant of 0.3 ms (3-dB bandwidth of ~230 Hz) and then averaging similarly to Ref. [25].

The sensor parameters (pump and probe power to the sensor, vapor cell temperature, field modulation amplitude) were optimized to provide high sensitivity at the target bandwidth of about 100 Hz. Before operation, the light shift and the residual magnetic field along the $z$-axis were zeroed. The sensor was calibrated using the shield coils with a 2-Hz square wave waveform with an amplitude of about ±200 pT. The sensor bandwidth was measured by applying a chirped waveform up to 400 Hz along the sensor sensitive axis using the shield coils; the chirp was generated using a dynamic signal analyzer (SR785, Stanford Research Systems).

### 4.2 *Sensor performance results*

Here, we report the detailed performance of a single sensor while in the next section we describe the performances of the 18 modules built so far. The vapor cell temperature of the sensor was around 135 °C and the pump and probe powers were set at 3.3 mW and 3.1 mW, respectively. The power ratio of the two pump beams was 27%. The 1-kHz modulation amplitude, optimized to give the highest signal, was about 140 nT.

Figs. 7A–B show the magnetic and gradiometrically inferred sensitivities of the four channels of the sensor. The gradiometrically inferred sensitivity is formed by first subtracting the signals of adjacent magnetometer channels and calculating the amplitude-spectral density of the difference signal. The amplitude-spectral density is further divided by $2^{1/2}$ to yield an equivalent single-channel sensitivity. The magnetic sensitivities of the four channels as averaged over 10–44 Hz range from 10.5 to 12.5 fT/rt-Hz while the corresponding gradiometrically inferred sensitivities range from 4.1 to 5.6 fT/rt-Hz, respectively. The respective gradiometer sensitivities range from 3.2 to 4.4 fT/(cm×rt-Hz).

Figs. 7C–D show the detailed noise contributions to the magnetic and gradiometrically inferred sensitivities for single magnetometer and gradiometer channels. For the shown magnetometer channel, the estimated pump-blocked (no pump light going to the sensor), photon-shot-noise, and electronic noise contributions are 3.3, 1.7 and 1.4 fT/rt-Hz, respectively.

The sensitivity of the magnetically insensitive quadrature component of the signal is 5.3 fT/rt-Hz. The corresponding gradiometrically inferred sensitivities are 3.3, 1.7 and 1.4 fT/rt-Hz, respectively, while the gradiometrically insensitive quadrature component has a sensitivity of 3.0 fT/rt-Hz. At higher frequencies, the gradiometer channel is operating close to the pump-blocked sensitivity with contributions from photon-shot, electronic and probe-beam noise: near 60 Hz, the gradiometrically inferred sensitivity is 3.6 fT/rt-Hz. The electronic noise level is limited by the resolution of the 18-bit analog-to-digital converter (ADC) of the DAQ. In the future MEG system, we plan to use a higher resolution ADC with a matched input range to the photodiode signal so that the electronic noise will be reduced. By doubling or tripling the probe power, the pump-blocked noise level could be further reduced by about 0.5–1 fT/rt-Hz, but there was not a significant effect on the magnetometer and gradiometer noise levels. From the vapor-cell temperature of 135 °C, we calculate the $^{87}$Rb atom density to be $\sim 5\times 10^{13}$/cm$^3$, which together with the sensor operating parameters give us an estimate of ~1.0 fT/rt-Hz for the atom-shot noise.

Figs. 8A and 8B show the frequency responses of the magnetometer and gradiometer channels, respectively. The estimated 3-dB atom bandwidths (see Eq. 5 in Ref. [25]) are 101, 120, 115 and 105 Hz for the four magnetometer channels, respectively. The low-frequency CMRRs of the gradiometer channels are approximately 100. At higher frequencies the CMRR decreases to about 20 due to differences in the 3-dB points of the magnetometer channels and resulting phase shifts. More uniform CMRR could likely be achieved by better magnetic field nulling (especially of $B_z$) at the channel locations. Fig. 8C shows the raw photodiode outputs of the four channels. Due to the ability to individually adjust the channel waveplates in this new sensor design, the signal voltage offsets are approximately zero.

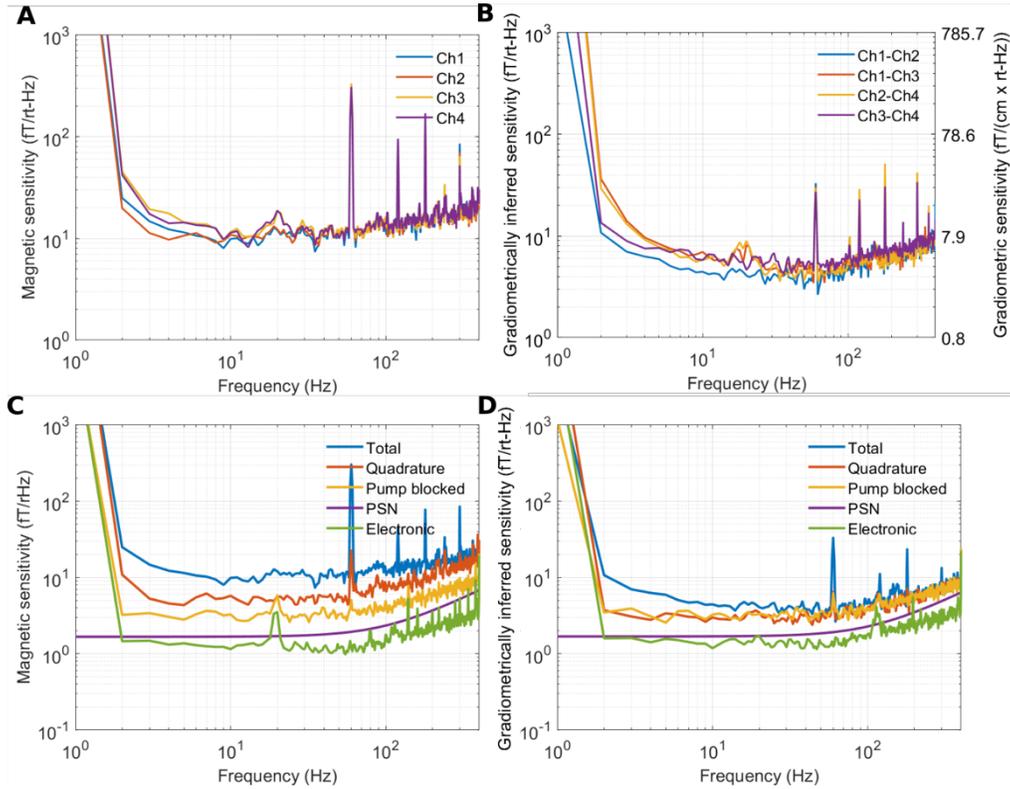

**Figure 7:** The bandwidth-normalized sensitivity of the sensor. **A:** The sensitivities of all four magnetometer channels of the sensor. **B**: The gradiometrically inferred and gradiometer sensitivities of the four gradiometer channels of the sensor. **C** and **D**: Detailed noise contributions to the magnetic (**C**) and gradiometrically inferred (**D**) sensitivities of single channels. The magnetically insensitive quadrature output of the sensor is formed by applying a 90° phase shift in the lock-in detection with respect to the magnetically-sensitive quadrature. The probe noise contribution is estimated by blocking the pump beam entering the sensor head. Photon-shot-noise (PSN) is estimated from the photocurrents at the two photodiodes and electronic noise is measured by blocking all light entering the sensor.

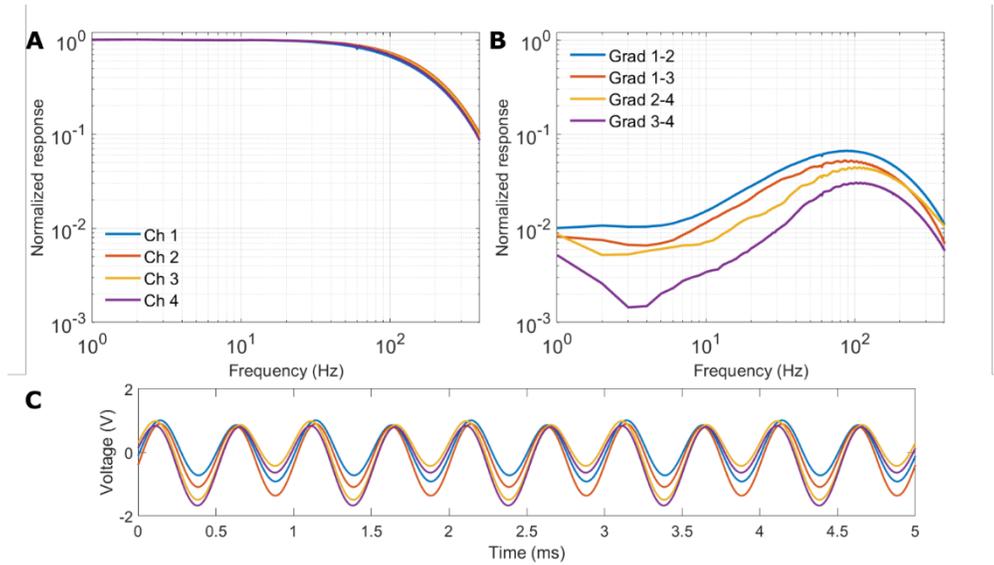

**Figure 8**: The frequency responses of the four magnetometer (**A**) and gradiometer (**B**) channels. **C:** The raw photodiode voltages of the four channels of the magnetometer.

### 4.3 *Sensor performance histograms*

Fig. 9 shows the histograms of the pump and probe power across the 18 built sensor modules as well as histograms of the magnetometer signal responses (V/nT), magnetic and gradiometrically inferred sensitivities as averaged over 10–44 Hz as well as atomic bandwidths across all the 4×18 channels. The sensors need on average 3.1 mW of pump power (median: 3.1 mW; range: 2.1–3.9 mW) and 3.1 mW of probe power (median: 3 mW; range: 2.96–4.0 mW) for optimal performance. The average magnetic sensitivity across the 72 channels is 12.3 fT/rt-Hz (standard deviation (SD): 1.4 fT/rt-Hz; range: 9.6–16.1 fT/rt-Hz) while the average gradiometrically inferred sensitivity is 6.0 fT/rt-Hz (SD: 1.5 fT/rt-Hz; range: 3.8–12.3 fT/rt-Hz). The bandwidth across the channels is on average 100 Hz, while its standard deviation and range are 12.4 Hz and 74–133 Hz, respectively.

In addition to the performance, we have quantified the remanent magnetic field differences between the channels within a sensor across twelve of the built sensor modules. We estimate that the absolute difference of the remanent transverse field ($B_x$ or $B_y$) between the channels is on average 0.4 nT and at maximum 2.9 nT within a sensor. The absolute difference of the longitudinal field ($B_z$) is on average 0.3 nT and at maximum 1.2 nT. The gradients inside the sensor together with the gradient inside the magnetic shield partly explain the channel bandwidth differences within the sensor and could be alleviated with individual $B_z$ control at the channels as implemented in the on-sensor coil design, but this is not used in these characterization measurements.

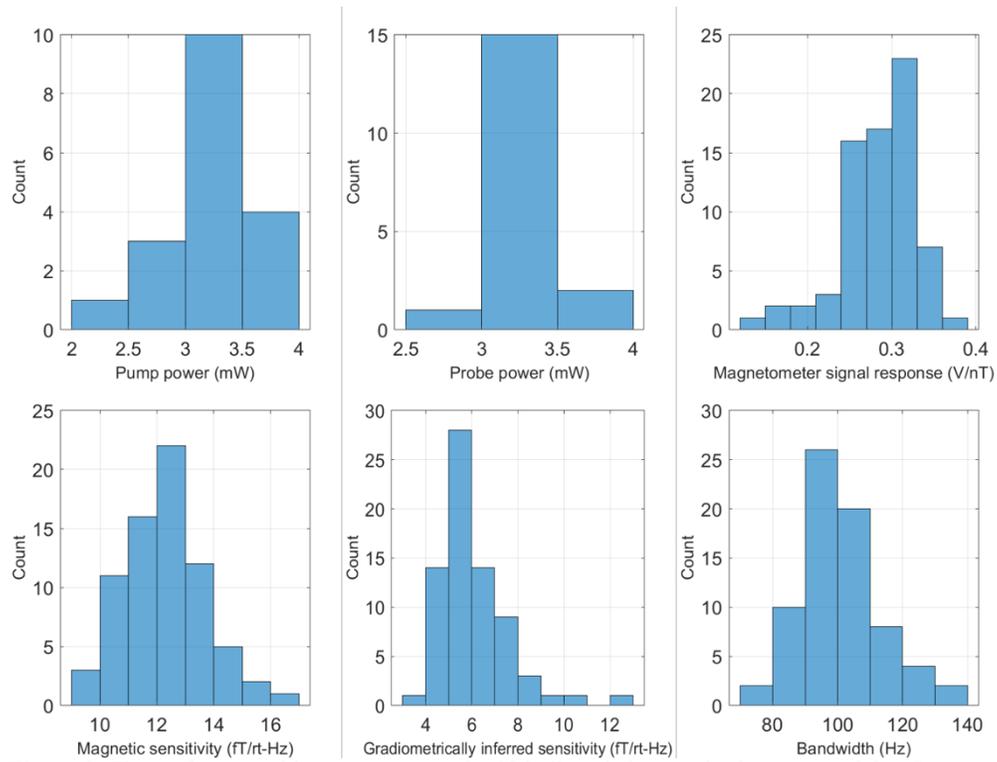
**Figure 9:** The performance histograms across the 18 built and characterized sensor modules. Pump and probe power are shown across the modules while the magnetometer-signal response (V/nT), magnetic and gradiometrically inferred sensitivity (averaged over 10–44 Hz) as well as atomic bandwidth are shown across the 4×18 channels.

## 5. Conclusions

We have presented our next-generation OPM sensor and outlined the performance of 18 sensor modules that we have built and characterized thus far. We have also presented a new on-sensor coil design based on stream-function optimization using an open-source software package *bfieldtools*. The magnetic (gradiometrically inferred) sensitivity of the sensor is on average 12.3 fT/rt-Hz (6.0 fT/rt-Hz). The highest sensitivity that the sensor achieves is 9.6 fT/rt-Hz (3.8 fT/rt-Hz). The sensor 3-dB bandwidth is approximately 100 Hz. We are currently in the process of building up nine additional sensor modules and implementing them into a 27-sensor (108-channel) whole-head MEG sensor array operating in a large magnetically shielded room. The optical pump and probe power will be supplied to the sensors using common pump and probe lasers distributed and combined using custom optical boards and fibers.

The new sensor design implements several improvements over our older sensor. First, the operating temperature of the vapor cell is reduced from roughly 150 °C to 135 °C. Second, the new detection optics design allows more uniform gradiometer sensitivity across the four channels of the sensor and reduced optical cross-talk between the channels. Third, we have substantially reduced the optical power required in the sensor operation. Our previous sensor used about 36 mW and 13 mW of pump and probe power, respectively, while the new sensor uses about 3 mW of both. This reduction in the optical power is beneficial in the multi-sensor operation using common seed lasers; it lowers the power requirements of the lasers and allows use of more sensors with a single laser. Fourth, the sensor has a new on-sensor coil design that allows control of the magnetic field along the optical axis at each channel location. This will be important in reducing the cross-axis projection errors in the sensor due to residual fields along the laser beams.

We did not stabilize the power of the pump and probe lasers in this work. However, we intend to power stabilize the lasers used in the MEG array operation. In that case it might be beneficial to increase the probe power for increased signal size and for reduced pump-blocked noise level. These factors together with the use of a high-resolution ADC and employment of the sensor in a large shielded room with potentially reduced magnetic Johnson–Nyquist and thermal magnetization noise compared to that in our table-top shield may further improve the magnetic and gradiometer sensitivity the sensor demonstrates.

In conclusion, we have developed a next-generation OPM sensor for MEG that is improved from our previous sensor. Our recent results suggest that our previous OPM sensor can record high-quality MEG data: an array of six such sensors (24 channels) achieved similar or slightly higher performance than SQUID sensors when classifying single-trial auditory evoked responses [45]. We thus expect that our new sensor implemented to a whole-head sensor array will similarly allow recording of high-quality MEG signals with the information conveyed by the OPM array even surpassing that of the previous array.

**Funding.** Research reported in this publication was supported by the Laboratory Directed Research and Development program at Sandia National Laboratories and the National Institute of Biomedical Imaging and Bioengineering of the National Institutes of Health, under Award Number U01EB028656.

**Acknowledgments.** The authors thank Drs. Bethany J. Little and Pauli Kehayias for their help with the sensor. The authors also want to thank Drs. Julia M. Stephen and Samu Taulu as well as Mr. Jim McKay for their collaboration in the magnetoencephalography project.

Sandia National Laboratories is a multimission laboratory managed and operated by National Technology & Engineering Solutions of Sandia, LLC, a wholly owned subsidiary of Honeywell International Inc., for the US Department of Energy National Nuclear Security Administration under contract DENA0003525. The employee, not NTESS, owns the right, title, and interest in and to the written work and is responsible for its contents. Any subjective views or opinions that might be expressed in the written work do not necessarily represent the views of the Department of Energy, the National Institutes of Health, or the United States Government. The publisher acknowledges that the U.S. Government retains a non-exclusive, paid-up, irrevocable, world-wide license to publish or reproduce the published form of this written work or allow others to do so, for U.S. Government purposes. The DOE will provide public access to results of federally sponsored research in accordance with the DOE Public Access Plan.

**Disclosures.** No disclosures.

**Data availability.** The datasets generated during and/or analyzed during the current study are available from the corresponding author on reasonable request.

**Supplemental document.** No supplementary material.